\documentclass[a4paper,superscriptaddress,prd,showpacs,preprintnumbers,10pt,eqsecnum,reprint]{revtex4}
\pdfoutput=1
\usepackage{amsmath,color}
\usepackage{color}
\usepackage{amsmath,amsfonts,amssymb,mathrsfs}
\usepackage{booktabs}
\usepackage{graphicx}
\usepackage{float} 	
\usepackage{array}
\usepackage{sepfootnotes}

\usepackage{longtable}
\allowdisplaybreaks
\AtBeginDocument{
	\heavyrulewidth=.08em
	\lightrulewidth=.05em
	\cmidrulewidth=.03em
	\belowrulesep=.65ex
	\belowbottomsep=0pt
	\aboverulesep=.4ex
	\abovetopsep=0pt
	\cmidrulesep=\doublerulesep
	\cmidrulekern=.5em
	\defaultaddspace=.5em
}

\usepackage{hyperref}								
\hypersetup{bookmarksopen=false,colorlinks,pdfstartview=FitH, citecolor=black,runcolor=black,anchorcolor=black,linkcolor=black,
	pdftitle = {PaperTitle}
}

\begin{document}
	
	\title{A vector meson dominance model for pions}

	\author{G. B. Tupper}
	\affiliation{Centre for Theoretical \& Mathematical Physics, and Department of Physics, University of
		Cape Town, Rondebosch 7700, South Africa.\\National Insitute for Theoretical Physics, Private Bag X1, Matieland 07602, South Africa.}
	
	\date{\today}
\maketitle

\indent    A renormalizable non-abelian vector meson dominance model coupled to pions is constructed which exhibits a ${\rm SO}\left(4,3\right)$ vacuum symmetry. Tree-level predictions for $p$-$p$ scattering lengths and slopes  at leading order are in reasonable agreement with experiment.
\bigskip

\indent    Vector meson dominance (VMD) dates back to Sakurai (1) who appIDd Yang-Mills theory to strong interactions to great success. An important step was taken by Kroll, Lee and Zumino (KLZ) (2) who showed how to make VMD for electromagnetic form factors compatible with gauge invariance. With the advent of the standard model ([3]) VMD has largely been subsumed into predicting the low energy constants of the chiral perturbation theory (CPTH) expansion [4].,
\bigskip

\indent    In more recent years a KLZ restricted to charged pion and the neutral rho has been used in studies of finite temperature (5) and form factors (6,7,8). On the other hand, the charged and neutral pion form an isospin triplet, as do the neutral and charged rho. In Sakurai's vision the vectors would be governed by a Yang-Mills theory. Conventional wisdom (4) provides this via massive Yang-Mills theory\footnote{Back in the labour preceding the birth of the Standard Model, Weinberg (9) noted chiral symmetry and VMD could be reconciled but his non-linear sigma model solution (which was reincarnated as Hidden Local Symmetry (10)) is not renormalizable.}. 
\bigskip

\noindent    In his seminal paper (11) 't Hooft exhibited a model with isospin and VMD but no matter. Previous work (12) without electromagnetism examined the question of what happens when pion matter is included in a similar model. In this work I address the issues of how to include electromagnetism in a model which is stable.
\bigskip

\indent   Inclusion of electromagnetism is simply a matter of adapting 't Hooft's (11) ``hyper-electromagnetism'': assign the Higgs doublet $\Phi $ a $U\left(1\right)$ ``hypercharge'' $Y\left(\Phi \right)=1$ and pions $Y\left(\vec{\pi }\right)=0$. The covariant derivatives are $D_{\mu } \vec{\pi }=\left(\partial _{\mu } +ig\vec{T}\cdot \vec{V}_{\mu } \right)\vec{\pi }$,  $D_{\mu } \Phi =\left(\partial _{\mu } +ig\vec{T}\cdot\vec{V}_{\mu } -e'B_{\mu } \right)\Phi $ Then a general model is
\begin{equation} \label{ZEqnNum529081} 
\mathcal{L} = {\textstyle\frac{1}{2}} \left(D_{\mu } \vec{\pi }\right)\left(D^{\mu } \vec{\pi }\right)-{\textstyle\frac{1}{2}} m_{\pi } ^{2} \vec{\pi }^{2} -{\textstyle\frac{1}{4}} \vec{V}_{\mu \nu } \vec{V}^{\mu \nu } -{\textstyle\frac{1}{4}} B_{\mu \nu } B^{\mu \nu } +\left(D^{\mu } \Phi \right)^{\dag } \left(D_{\mu } \Phi \right)-V\left(\left|\Phi \right|,\left|\vec{\pi }\right|\right) 
\end{equation} 
 In the unitary gauge (13) with $\mu ^{2} =\lambda 'v^{2} $, $\Phi ^{T} ={\left(0,v+\sigma \right)\mathord{\left/ {\vphantom {\left(0,v+\sigma \right) \sqrt{2} }} \right. \kern-\nulldelimiterspace} \sqrt{2} } $, redefinitions $e={e'g\mathord{\left/ {\vphantom {e'g \sqrt{g^{2} +e'^{2} } }} \right. \kern-\nulldelimiterspace} \sqrt{g^{2} +e'^{2} } } $

\noindent $V_{a} ^{\mu } =\rho _{a} ^{\mu } +\left({e'\mathord{\left/ {\vphantom {e' g}} \right. \kern-\nulldelimiterspace} g} \right)B^{\mu } \delta _{a3} $,$A_{\mu } =\sqrt{1+{e'^{2} \mathord{\left/ {\vphantom {e'^{2}  g^{2} }} \right. \kern-\nulldelimiterspace} g^{2} } } B_{\mu } $  yield
\begin{align} \label{ZEqnNum536895} 
\mathcal{L}_{U\, {\rm gauge}} &= \frac{1}{2} \left(D_{\mu } \vec{\pi }\right)\left(D^{\mu } \vec{\pi }\right)-{\textstyle\frac{1}{4}} \vec{\rho }_{\mu \nu } \cdot \vec{\rho }^{\mu \nu } -{\textstyle\frac{1}{4}} F_{\mu \nu } F^{\mu\nu } -\frac{e}{2g} F_{\mu \nu } \rho _{3} ^{\mu \nu } +{\textstyle\frac{1}{2}} \left(\partial^{\mu } \sigma \right)\left(\partial_{\mu } \sigma \right) +\nonumber \\ 
&-V\left(\left|\Phi \right|,\left|\vec{\pi }\right|\right)+\frac{g^{2} }{8} \left(v+\sigma \right)^{2} V_{a}^{\mu } V_{a\mu} ,\quad D_{\mu } \vec{\pi }=\left(\partial _{\mu } +ig\vec{T}\cdot\vec{V}_{\mu } +ieT_{3} A_{\mu } \right)\vec{\pi }
\end{align} 
The potential must have a minimum$V\left({v\mathord{\left/ {\vphantom {v 2}} \right. \kern-\nulldelimiterspace} 2} ,0\right)$ and the most general renormalizable one is
\begin{equation} \label{ZEqnNum786257} 
V\left(\left|\Phi \right|,\left|\vec{\pi }\right|\right)=\lambda \left|\Phi ^{\dag } \Phi \right|^{2} -\lambda v^{2} \left|\Phi ^{\dag } \Phi \right|+\frac{\lambda '}{8} \left(\vec{\pi }^{2} \right)^{2} +\frac{\kappa }{2} \vec{\pi }^{2} \left(\left|\Phi ^{\dag } \Phi \right|-v^{2} \right) 
\end{equation} 
\bigskip

\noindent The  Higgs mass is $m_{\sigma } =\sqrt{2\lambda } v$ and $m_{\rho } ={gv\mathord{\left/ {\vphantom {gv 2}} \right. \kern-\nulldelimiterspace} 2} $ Due to a mixing angle $\sin \vartheta ={eg\mathord{\left/ {\vphantom {eg \sqrt{g^{2} +e^{2} } }} \right. \kern-\nulldelimiterspace} \sqrt{g^{2} +e^{2} } } $  $m_{\rho ^{\pm } } ={gv\mathord{\left/ {\vphantom {gv 2}} \right. \kern-\nulldelimiterspace} 2} ,m_{\rho ^{0} } ={\sqrt{g^{2} +e^{2} } v\mathord{\left/ {\vphantom {\sqrt{g^{2} +e^{2} } v 2}} \right. \kern-\nulldelimiterspace} 2} $ and$m_{\gamma } =0$ .Note the last term in (\ref{ZEqnNum786257}), which is required by renormalizability,  connects the scalars in$\left(\Phi ,\vec{\pi }\right)$ space. It has been arranged so that fluctuations about the minimum do not contribute to the ``pion mass'' appearing in (\ref{ZEqnNum529081}). That is a key departure from [12]; one can just as well write for a scalar source$J$ later set to$m_{\pi } ^{2} $ 
\begin{equation} \label{ZEqnNum800567} 
\mathcal{L}={\textstyle\frac{1}{2}} \left(D_{\mu } \vec{\pi }\right)\left(D^{\mu } \vec{\pi }\right)-{\textstyle\frac{1}{2}} J\vec{\pi }^{2} -{\textstyle\frac{1}{4}} \vec{V}_{\mu \nu }\cdot \vec{V}^{\mu \nu } -{\textstyle\frac{1}{4}} B_{\mu \nu } B^{\mu \nu } +\left(D^{\mu } \Phi \right)^{\dag } \left(D_{\mu } \Phi \right)-V\left(\left|\Phi \right|,\left|\vec{\pi }\right|\right) 
\end{equation} 
In that way one makes contact with CHPT and the pion scalar form factor. In essence one has a  $SO\left(4)\otimes SO\right(3)$ linear sigma model treating massless pions and would-be Nambu-Goldstone bosons that are ``eaten'' by the vector on the same footing.
\bigskip

\indent    Now the $p$-$p$ scattering amplitude is easily calculated and most compactly expressed
\begin{equation} \label{ZEqnNum415080} 
\begin{array}{c} {T\left(s,t,u\right)=A\left(s,t,u\right)\delta _{ab} \delta _{cd} +A\left(t,s,u\right)\delta _{ac} \delta _{bd} +A\left(u,t,s\right)\delta _{ad} \delta _{bc} } \\ {A\left(s,t,u\right)=-\lambda '+\frac{\left(\kappa v\right)^{2} }{m_{\sigma } ^{2} -s} +\frac{g^{2} \left(s-u\right)}{m_{\rho } ^{2} -t} +\frac{g^{2} \left(s-t\right)}{m_{\rho } ^{2} -u} } \end{array} 
\end{equation} 
The first two terms in$A\left(s,t,u\right)$ are present in any scalar sigma model, While the scalar couplings in  (\ref{ZEqnNum786257}) are a-priori independent, a constraint$\kappa ^{2} =2\lambda \lambda '$ causes  $p$-$p$ scattering amplitude (\ref{ZEqnNum415080}) to vanish at threshold for massless pions
\begin{equation} \label{6)} 
A\left(s,t,u\right)=\frac{\lambda 's}{m_{\sigma } ^{2} -s} +\frac{g^{2} \left(s-u\right)}{m_{\rho } ^{2} -t} +\frac{g^{2} \left(s-t\right)}{m_{\rho } ^{2} -u}  
\end{equation} 
It should be stressed that for $m_{\pi } =0$ the condition $A\left(0,0,0\right)=0$ is connected  in Weinberg's classic analysis (14)  to derivative coupling (as for the rho -exchange pieces), but there are no additional inputs from current algebra or PCAC here. Defining a scale factor $a\equiv \sqrt[{4}]{{\lambda \mathord{\left/ {\vphantom {\lambda  2\lambda '}} \right. \kern-\nulldelimiterspace} 2\lambda '} } $one has a unique potential\footnote{  The potential in (12) has   $a^{2} <0$   and is  unstable to  $\left\langle \Phi \right\rangle \ne 0 , \left\langle \vec{\pi }\right\rangle \ne 0$ away from the chiral limit. } obeying the $A\left(0,0,0\right)=0$constraint :
\begin{equation} \label{ZEqnNum342488} 
V=\frac{\lambda }{4} \left(\Phi ^{\dag } \Phi -a^{2} \, \vec{\pi }^{2} -\frac{v^{2} }{2} \right)^{2}  
\end{equation} 
The vacuum group structure is ${\rm SO}\left(4,3\right)$. 
\bigskip

\indent Relevant parameters are the couplings$g,\lambda $  and scale factor.  In these terms the scattering lengths and slopes [12] can be reorganised into an  ${m_{\pi } \mathord{\left/ {\vphantom {m_{\pi }  v}} \right. \kern-\nulldelimiterspace} v} $ expansion with leading terms independent of the Higgs self-coupling
\begin{equation} \label{8)} 
 .\begin{array}{c} {\begin{array}{cc} {a_{0}^{0} =\frac{\left[16+a^{4} \right]m_{\pi } ^{2} }{8\pi v^{2} } } & {b_{0}^{0} =} \end{array}\frac{\left[8+a^{4} \right]m_{\pi } ^{2} }{16\pi v^{2} } } \\ {\begin{array}{ccc} {a_{1}^{1} =\frac{\left[12+a^{4} \right]m_{\pi } ^{2} }{24\pi v^{2} } } & {a_{0}^{2} =-\frac{m_{\pi } ^{2} }{\pi v^{2} } } & {b_{0}^{2} =-} \end{array}\frac{\left[8+a^{4} \right]m_{\pi } ^{2} }{8\pi v^{2} } } \end{array}.   
\end{equation} 
   The natural value is $a=\sqrt{{\pi \mathord{\left/ {\vphantom {\pi  2}} \right. \kern-\nulldelimiterspace} 2} } $ which makes the vacuum 3-sphere and 4-sphere areas equal. One cannot impose the KSRF (15) relation because it draws upon$\rho ^{0} \to e^{+} e^{-} $ at a scale$m_{\rho } $ whereas the appropriate scale in $p$-$p$ scattering is$m_{\pi } $.  One infers a running gauge coupling driven by kinematic dominant pion loops and, modifying the expressions in [5] by a factor of ${3\mathord{\left/ {\vphantom {3 2}} \right. \kern-\nulldelimiterspace} 2} $ since $\pi ^{0} $ also contributes, the leading log is ${1\mathord{\left/ {\vphantom {1 g^{2} _{{\rm VMD}} }} \right. \kern-\nulldelimiterspace} g^{2} _{{\rm VMD}} } ={1\mathord{\left/ {\vphantom {1 g^{2} _{{\rm KSRF}} +{\ln \left({m_{\rho } \mathord{\left/ {\vphantom {m_{\rho }  m_{\pi } }} \right. \kern-\nulldelimiterspace} m_{\pi } } \right)\mathord{\left/ {\vphantom {\ln \left({m_{\rho } \mathord{\left/ {\vphantom {m_{\rho }  m_{\pi } }} \right. \kern-\nulldelimiterspace} m_{\pi } } \right) 12\pi ^{2} }} \right. \kern-\nulldelimiterspace} 12\pi ^{2} } }} \right. \kern-\nulldelimiterspace} g^{2} _{{\rm KSRF}} +{\ln \left({m_{\rho } \mathord{\left/ {\vphantom {m_{\rho }  m_{\pi } }} \right. \kern-\nulldelimiterspace} m_{\pi } } \right)\mathord{\left/ {\vphantom {\ln \left({m_{\rho } \mathord{\left/ {\vphantom {m_{\rho }  m_{\pi } }} \right. \kern-\nulldelimiterspace} m_{\pi } } \right) 12\pi ^{2} }} \right. \kern-\nulldelimiterspace} 12\pi ^{2} } } $. Taking the VMD value $g=g_{\rho \pi \pi } =5$ gives $v=308MeV$, so the  Higgs mass is $m_{\sigma } =436\sqrt{\lambda } {\rm MeV}$. The predictions are $a_{0}^{0} =0.145\, \left[0.22\right]$ , $b_{0}^{0} =0.041\, \left[0.25\right]$ , $a_{1}^{1} =0.038\; [0.038]$,$a_{0}^{2} =-0.063\, \left[-0.044\right]$ and $b_{0}^{2} =-0.082\, \left[-0.082\right]$ ; note $a_{0}^{0} -5a_{0}^{2} =0.46\, \left[0.44\right]$  is independent of the condition and in much better agreement than the individual lengths. What is striking is precise agreement in the $I=J=1$ channel where VMD should hold. Overall the (dis)agreement is as good as one should expect of any tree-level model.
\bigskip
\noindent     

\noindent   One now has a VMD model which is renormalizable, non-abelian and illustrates the degree to which vector meson dominance is an alternative to chiral perturbation theory. One-loop calculations are facilitated in the unitary background field gauge (16) wherein the unitary gauge is applied to the background fields while maintaining a renormalizable 't Hooft gauge on loops and offering simple Ward identities. It can be coupled to nucleons as external sources [1] but for pseudo-vector current renormalizability is lost. Ceding that point\footnote{ex quo in eo loco positum sit}, its vacuum group structure can be promoted to a gauged ${\rm SO}\left(4,3\right)$non-linear sigma model, but such things are beyond the present scope.

\noindent 

\noindent \textbf{}

\noindent 
\subsection{Acknowledgements}

\noindent I wish to thank C. Dominguez, P. Moodley and K. Schilcher for discussions on KLZ. This work was supported by the National Research Foundation (South Africa).

\noindent 

\noindent 
\subsection{References}

\begin{enumerate}
\item  J. J. Sakurai, Ann. Phys. (N.Y.) \textbf{11}, 1 (1960). https://doi.org/10.1016/0003-4916(60)90126-3 

\item  N. M. Kroll, T. D. Lee, and B. Zumino, Phys. Rev. \textbf{157}, 1376 (1967).

\item  J. F Donoghue, E. Golowich and B. R Holstein, \textit{Dynamics of the Standard Model}, Cambridge University Press (2014).

\item   J. F. Donoghue, C. Ramirez and G. Valencia, Phys. Rev. \textbf{D39}, 1947 (1989)

\item  C. Gale and J. Kapusta, Nucl. Phys. \textbf{B357}, 65 (1991).

\item  C.A. Dominguez, M. Loewe, J.I. Jottar, and B. Willers, Phys. Rev. \textbf{D76}, 095002 (2007).

\item  C. A. Dominguez, M. Loewe, and B. Willers, Phys. Rev. \textbf{D 78}, 057901 (2008).

\item  C. A. Dominguez, M. Loewe, and M. Lushozi, Adv. High Energy Phys. 2015 (2015), 803232.

\item  S. Weinberg, Phys. Rev.\textbf{166}, 1568 (1968).

\item  M. Bando,T. Kugo, S. Uehara, K. Yamawaki and T. Yanagida, Phys. Rev. Lett. 54(1985) 1215.

\item  G. 't Hooft, Nucl. Phys. B3, 167 (1971).

\item  C. A. Dominguez, P. Moodley, K. Schilcher, and G. B. Tupper, arXiv:1711.07204v1

\item  A. Datta and P.. Ghose, Phys. Rev. \textbf{D 9}, 2465 (1974).

\item  S. Weinberg, Phys. Rev. Lett. 17, 616 (1966);18, 188 (1967).

\item  K. Kawarabayashi and M. Suzuki, Phys. Rev. Lett. 16,

\noindent 255 (1966); Riazuddin and Fayyazuddin, Phys. Rev. 147,1071 (1966)

\item  G. B. Tupper, arXiv: 1412.5959.
\end{enumerate}

\noindent

\end{document}